\documentclass[reprint,amsmath,amssymb,aps,prl]{revtex4-2}
\usepackage{CJK}
\usepackage{graphicx}
\usepackage{color}
\usepackage{amsmath}
\usepackage{amssymb}
\usepackage{amsthm}
\usepackage{extarrows}
\usepackage{tabularx}
\usepackage{dsfont}
\usepackage{booktabs}
\usepackage{float}
\usepackage{lineno}
\usepackage{csquotes}

\usepackage{appendix}

\usepackage{xcolor}

\usepackage[dvipdfm,colorlinks,urlcolor=blue,linkcolor=blue,anchorcolor=blue,citecolor=blue]{hyperref}

\begin{document}

\title{Gravity-Induced Thermal Rectification in Gaseous Systems}

\author{Rongxiang Luo$^{1}$}
\author{Chao Yang$^1$}
\author{Juncheng Guo$^1$}
\author{Qiyuan Zhang$^{2}$}


\affiliation{
\vspace{1em}
$^1$ Department of Physics, Fuzhou University, Fuzhou 350108, Fujian, China\\
$^2$ Department of Physics, Xiamen University, Xiamen 361005, Fujian, China
\vspace{1em}}
\date{\today }
\begin{abstract}
Thermal rectification (TR) typically relies on structural asymmetry or material heterogeneity. Here, we show that gravity alone can induce and modulate TR in gaseous systems. Using a minimal model of a single gas particle confined in a two-dimensional channel between heat baths at different temperatures, we analytically demonstrate that gravitational fields generate TR, enabling perfect unidirectional heat conduction across broad gravitational parameter ranges. This effect exhibits an intrinsic trade-off between rectification efficiency and heat power. Extending beyond the single-particle limit, numerical simulations confirm that gravitationally induced TR persists in interacting many-particle systems. Notably, in interacting gas mixtures, the rectification direction can be reversed. Gravity-mediated control of heat currents thus provides new fundamental insights into thermal transport and suggests design principles for gaseous thermal diodes.
\end{abstract}

\maketitle

\textit{Introduction}---The precise control of heat current is a central goal in modern thermal physics, driving the pursuit of novel mechanisms for thermal management and energy conversion~\cite{Li2012,BENENTI20171,Huang2024}. A paradigmatic example is thermal rectification (TR), the thermal analog of an electrical diode, in which heat transport is favored in one direction. To date, diverse mechanisms involving phonons~\cite{Terraneo2002,Li2004,Donovan2020}, electrons~\cite{Segal2008}, and photons~\cite{Otey2010} have been investigated across classical and quantum frameworks, with comprehensive reviews in Refs.~\cite{Roberts2011,Wehmeyer2017,Benenti2023}. In these systems, TR typically arises from intrinsic nonlinearity and structural asymmetry, with substantial experimental efforts devoted to validating these approaches~\cite{Chang2006,Wang2017,Yang2020,Yufeng2022}. Despite these advances, TR in conventional gaseous media---particularly the roles of external forces and interparticle interactions---remains largely unexplored.\par

Realizing TR in gases poses a fundamental challenge. Unlike crystalline solids, where atoms are anchored by interatomic bonds, strategies effective in solid-state systems---such as exploiting material interfaces~\cite{Li2005,Riccardo2018}, mass gradients~\cite{Dettori2016,Wang2012,Pereira2010,Yang2007}, or temperature-dependent properties~\cite{Jiping2015,Kobayashi2009}---are impractical in gaseous systems. This limitation stems from the intrinsic symmetry of particle interactions and the natural tendency of gases toward homogeneous equilibration. Efforts to break thermal reciprocity in gases have explored electron transport under magnetic fields ~\cite{Casati2007} and, more recently, ballistic conduction in geometrically asymmetric channels~\cite{2024zou,Yuanchen2021}. However, a clear mechanism for TR in conventional gaseous media, without relying on elaborate structural engineering, remains elusive.\par

Gravity, as a ubiquitous external field, profoundly influences the thermophysical properties of diverse gaseous systems~\cite{Barmatz2007,Weichman2001,Thorsten2001,van1985,RevModPhys.51.79}, from macromolecular and granular gases to simple fluids near their critical points. In macroscopic thermodynamics, gravity manifests primarily through height-dependent density distributions, which provide a natural source of spatial symmetry breaking conducive to TR. Within non-equilibrium thermodynamics, heat transport phenomena such as low-dimensional thermal conduction~\cite{lepri2016} and gravity-driven Rayleigh-B\'{e}nard convection~\cite{Straughan2024} are well described by continuum hydrodynamics. Yet this continuum assumption breaks down when gaseous systems are confined to microscale dimensions or subjected to low pressures ~\cite{Roohi2025,Rostami2002}. Consequently, the influence of gravity on ballistic or kinetic transport regimes presents a distinct avenue for manipulating heat current. These considerations motivate the fundamental questions about how gravity affects heat transport in non-continuum regimes, whether it can induce TR in gaseous systems, and the underlying mechanisms involved.\par

Here, we show that gravity alone is sufficient to induce robust TR. We first develop a minimal single-particle model to analytically establish that gravity enables perfect unidirectional conduction, revealing a critical tradeoff between rectification efficiency and heat power. We then numerically validate this mechanism in interacting many-particle systems. Remarkably, we find that in interacting gas mixtures, the rectification direction can reverse relative to the single-component case. By elucidating the mechanism of gravity-induced TR, our work suggests a route to thermal management in gaseous systems mediated by external fields.\par


\begin{figure}[t]
\centering
\includegraphics[width=6cm]{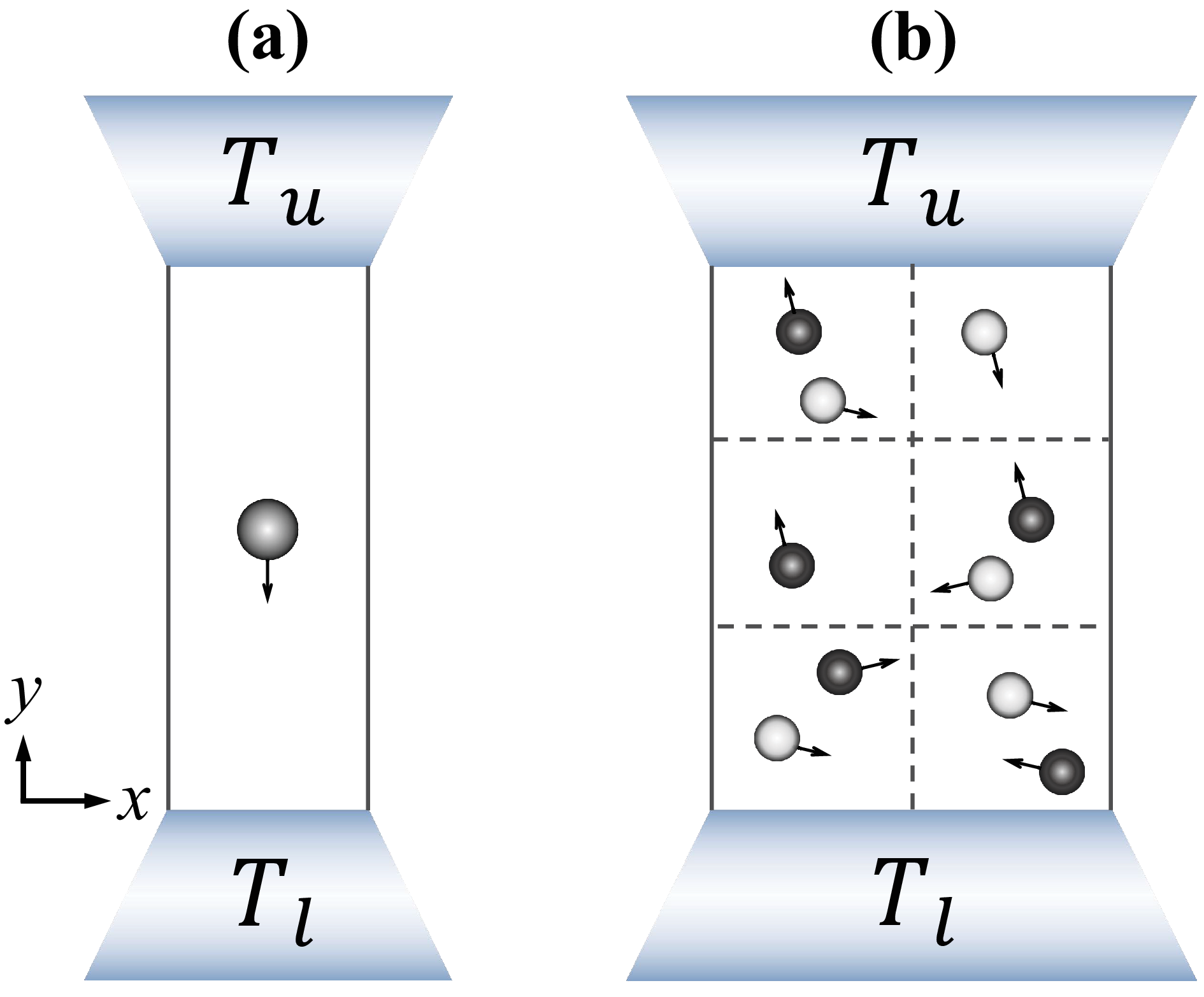}
\caption{Schematic illustration of the two-dimensional gaseous models under a gravitational field $\mathbf{g} = -g \hat{\mathbf{y}}$. (a) Single-particle system: a particle confined between heat baths maintained at temperatures $T_l$ (bottom, $y = 0$) and $T_u$ (top, $y = H$). (b) Interacting multi-particle mixture: a binary mixture of particles (distinguished by color/shading) undergoes collisions within a spatial grid (dashed lines) employed for the multi-particle collision dynamics method. Arrows indicate particle velocities.}
\label{model}
\end{figure}

\textit{Theoretical Model and Analytical Results}---As illustrated in Fig.~\ref{model}(a), we consider a minimal model of a point particle of mass $m$ confined within a two-dimensional rectangular channel of length $L$ and height $H$ in the $x-y$ plane, subject to a uniform gravitational field $\mathbf{g} = -g \hat{\mathbf{y}}$. The system is governed by the Hamiltonian
\begin{equation}\label{H}
  \mathcal{H}=\frac{p^2}{2m}+mgh, \qquad 0<h<H,
\end{equation}
with the upper and lower boundaries coupled to two heat baths at distinct temperatures. Each bath is modeled as a Maxwell reservoir~\cite{1978T,T1998}: when the particle collides with bath $\alpha$ $(\alpha = u,l)$, it is reflected with a random velocity $\mathbf{v} = (v_x, v_y)$, whose components are sampled from the respective distributions:
\begin{equation}\label{Eqfv}
\begin{aligned}
  f_{\alpha}(v_x) &= \sqrt{\frac{m}{2\pi k_{\mathrm{B}} T_\alpha}} \exp\!\left(-\frac{m v_x^2}{2k_{\mathrm{B}} T_\alpha}\right), \\
  f_{\alpha}(v_y) &= \frac{m |v_y|}{k_{\mathrm{B}} T_\alpha} \exp\!\left(-\frac{m v_y^2}{2k_{\mathrm{B}} T_\alpha}\right),
\end{aligned}
\end{equation}
where $T_\alpha$ denotes the temperature of bath $\alpha$ and $k_{\mathrm{B}}$ is the Boltzmann constant. The sampling enforces the physical constraint $v_y > 0$ for reflections from $y=0$ and $v_y < 0$ for reflections from $y=H$; $v_x$ remains unrestricted. Periodic boundary conditions are imposed along the $x$-direction. Through repeated interactions with the two baths, the system evolves toward a nonequilibrium steady state sustained by a persistent heat current. By isolating the essential physics of gravity-mediated heat transport, this setup provides a minimal prototype for gravitational TR and an analytical foundation for interpreting the more complex behavior observed in interacting systems.\par

Following the standard definition~\cite{Zou2025}, the heat current $J$ from the lower heat bath to the upper bath in our system is given by (see the Supplemental Material of Ref.~\cite{Zhang2026} for the derivation):
\begin{equation}\label{J}
  J = \frac{E_{l\to u}-E_{u\to l}}{t_{l\to u}+t_{u\to l}}=
 \frac{3k_{\mathrm{B}}}{2} \frac{T_l-T_u}{ t_{l\to u}+t_{u\to l}},
\end{equation}
where $E_{l\to u}$ and $t_{l\to u}$ denote, respectively, the average energy transferred by the particle and the average transit time from the lower bath to the upper bath; $E_{u\to l}$ and $t_{u\to l}$ are the corresponding quantities for the reverse direction. The explicit expressions for the transit times are
\begin{equation}\label{Eqtlu}
\begin{aligned}
  t_{l\rightarrow u} &= \frac{1}{g}\sqrt{\frac{\pi k_{\mathrm{B}} T_l}{2m}}\left\{e^{\frac{mgH}{k_{\mathrm{B}}T_l}}\left[1+\mathrm{erf}(\sqrt{\frac{mgH}{k_{\mathrm{B}}T_l}})\right]
  -1\right\}\\ &-\sqrt{\frac{2H}{g}},\\
  t_{u\rightarrow l} &= \frac{1}{g}\sqrt{\frac{\pi k_{\mathrm{B}} T_u}{2m}}\left\{e^{\frac{mgH}{k_{\mathrm{B}}T_u}}\left[1-\mathrm{erf}(\sqrt{\frac{mgH}{k_{\mathrm{B}}T_u}})\right]
  -1\right\}\\ &+\sqrt{\frac{2H}{g}}.
\end{aligned}
\end{equation}

We denote the temperatures of the hot and cold baths by $T_h$ and $T_c$, respectively. The \enquote{forward flux} $J_f$ is defined as the heat current from the hot lower bath ($T_l=T_h$) to the cold upper bath ($T_u=T_c$). Conversely, the \enquote{reverse flux} $J_r$ denotes the current from the hot upper bath ($T_u=T_h$) to the cold lower bath ($T_l=T_c$). Because only particles capable of overcoming the gravitational potential barrier contribute as effective energy carriers, gravity induces an asymmetry between the forward and reverse heat fluxes. With gravity acting along the negative $y$-direction, we expect $J_f>J_r$. In the forward configuration, a particle at the lower boundary absorbs thermal energy from the hot bath, increasing the probability that it attains a vertical velocity $v_y \geq \sqrt{2gH}$. This allows the particle to surmount the gravitational potential and reach the upper boundary, significantly enhancing the effective energy transport rate relative to the reverse configuration.\par

The mechanism underlying this TR effect can be understood by examining the round-trip rate $f$, defined as
\begin{equation}\label{f}
  f = \frac{1}{t_{l\to u} + t_{u\to l}}.
\end{equation}
From Eq.~(\ref{J}), it follows that $J\propto f$, implying that for fixed $T_l$ and $T_u$, TR emerges when the round-trip rate in the forward configuration ($f_f$) exceeds that in the reverse configuration ($f_r$). Following Ref.~\cite{2024zou}, we define the rectification efficiency $\xi$, which quantifies the directional asymmetry of heat transport, and the rectification power $P$ as:
\begin{equation}\label{xi}
 \xi =\frac{J_f-J_r}{J_f+J_r}, \quad \quad P=J_f-J_r.
\end{equation}
A nonzero $P$ signals the presence of TR, with larger magnitudes indicating a stronger rectification effect. The efficiency is bounded by $0\leq |\xi|\leq1$, where larger $|\xi|$ reflects more pronounced rectification, reaching perfect unidirectional conduction at  $|\xi|=1$ (corresponding to $J_r=0$ or $J_f=0$). Notably, negative values of $\xi$ and $P$ (indicating $J_r > J_f$) signify a reversed TR effect.\par

\begin{figure}
\centering
\includegraphics[width=8.5cm]{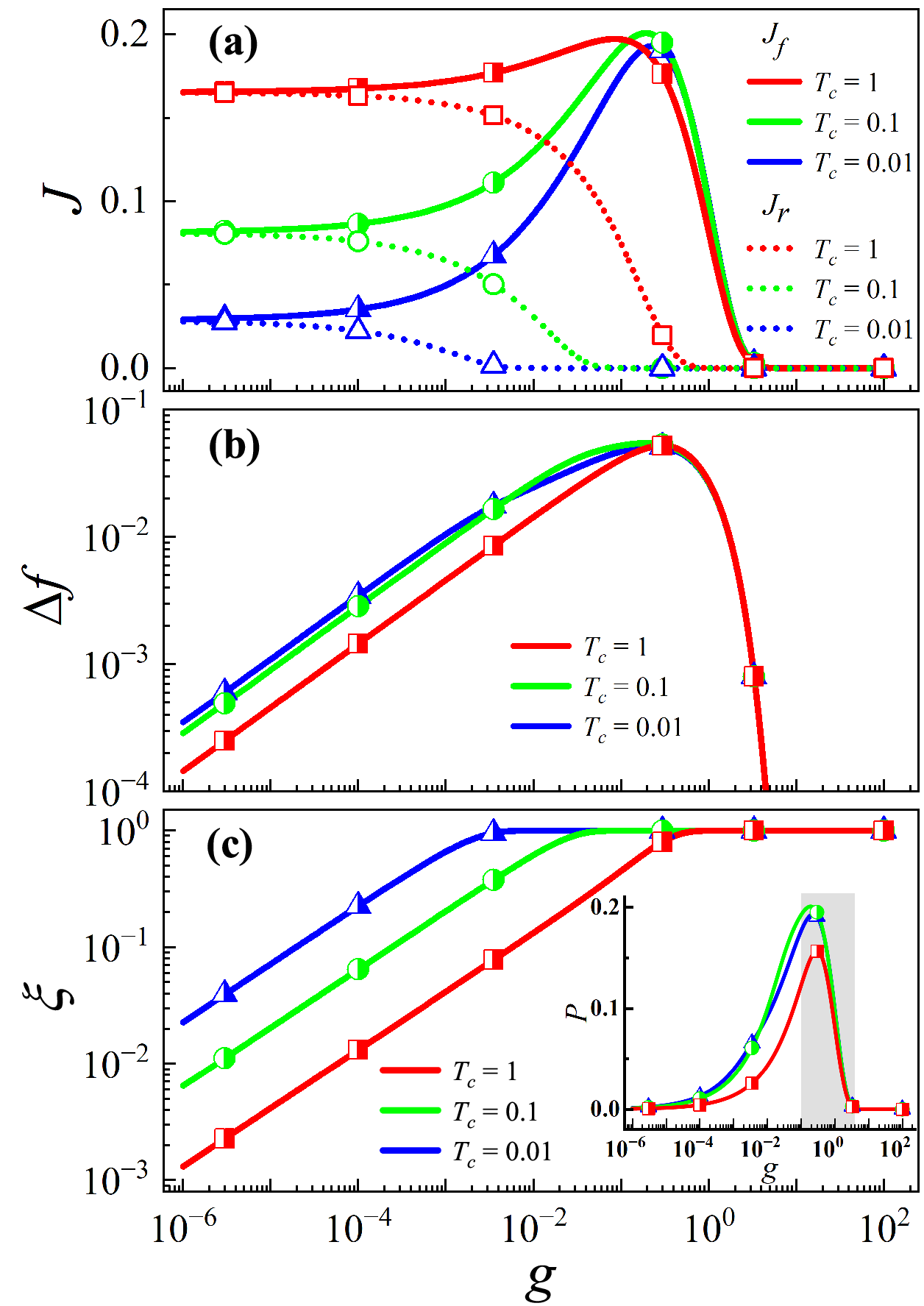}
\caption{(a) Heat current $J$, (b) round-trip rate difference $\Delta f$, and (c) rectification efficiency $\xi$ versus gravitational field strength $g$ for three cold-bath temperatures $T_c$. The colored curves in (a)--(c) represent analytical predictions from Eqs.~(\ref{J}), (\ref{f}), and (\ref{xi}), respectively; the accompanying symbols denote simulation data. Inset of (c): Power $P$ versus $g$. The gray shaded region ($g\in (0.1, 4)$) indicates the parameter regime where $\xi=1$  for both $T_c=0.1$ and $0.01$. Fixed parameters: $T_h=5$, $H=10$, $L=2$, and $m=k_B=1$.}
\label{fig2}
\end{figure}

Figure~\ref{fig2} presents our main theoretical results for the noninteracting gas. For a fixed hot-bath temperature $T_h=5$ and cold-bath temperatures $T_c=1$, $0.1$, and $0.01$ (distinct colored curves), the analytical predictions agree excellently with simulation data (symbols)~\cite{simulation}. Figure~\ref{fig2}(a) shows pronounced TR for all three $T_c$ values, with $J_f$ significantly exceeding $J_r$ over a broad range of $g$. As discussed, this effect originates from the difference in particle round-trip rates with the upper bath between the forward and reverse configurations. Figure~\ref{fig2}(b) plots the round-trip rate difference $\Delta f = f_f - f_r$. In agreement with our expectations, $\Delta f > 0$ throughout the rectification region for all tested $T_c$ values, peaking near $g\approx0.1$ before declining sharply at strong gravity.\par

Figure~\ref{fig2}(c) and its inset display the rectification efficiency $\xi$ and the power $P$, respectively. As $g$ increases, $\xi$ rises monotonically toward unity, corresponding to unidirectional heat flow, whereas $P$ exhibits a nonmonotonic behavior with a well-defined maximum $P_{\textrm{max}}$. This trade-off between power and efficiency is a central consideration in thermal diode design. The efficiency at maximum power can be obtained by solving $\partial P/\partial g=0$ for the critical point $g_\textrm{c}$ and substituting it into Eq.~(\ref{xi}). Notably, within the gray shaded region $g\in (0.1, 4)$, all nonzero $P$ values (including $P_{\textrm{max}}$) correspond to $\xi=1$ for both $T_c=0.1$ and $0.01$. This demonstrates that unidirectional heat flow can be maintained while the power is continuously tuned over a certain range of $g$. Moreover, lower $T_c$ accelerates the approach of $\xi$ to unity. Physically, a colder lower bath makes it increasingly difficult for a particle to overcome gravity and reach the upper hot bath. Consequently, as shown in Fig.~\ref{fig2}(a), $J_r$ decays rapidly with decreasing $T_c$, driving the system into a unidirectional conducting state. In the strong-gravity limit ($g>10$), where particle transport is effectively suppressed, both $J_f$ and $J_r$ vanish; although the mathematical limit yields $\xi=1$, we emphasize that practical relevance requires a finite $P$. Finally, in the weak-gravity region ($g<0.01$), while lower $T_c$ enhances $\xi$, the corresponding power curves for $T_c=0.1$ and $0.01$ remain nearly indistinguishable.\par

Interestingly, as shown in Fig.~\ref{fig2}(a), $J_f$ initially increases with $g$ before decaying, indicating that a weak gravitational field can enhance the effective thermal conductivity of the gaseous system in the forward configuration. This behavior arises from a velocity-filtering mechanism: when gravity opposes the thermodynamic force, a weak $g$ selectively admits high-energy particles, thereby enhancing inter-bath energy exchange. To illustrate this mechanism, consider an extreme scenario: if a particle acquires a near-zero velocity ($v_y \to 0$) after colliding with the hot lower bath, gravity rapidly drives it back for re-thermalization. This process accelerates particle recycling, thereby facilitating energy exchange. However, this mechanism does not apply to the reverse configuration; as illustrated in Fig.~\ref{fig2}(a), $J_r$ decreases monotonically with $g$, exhibiting no extremum where $\partial J_r/\partial g=0$. These results demonstrate that gravity-induced TR is highly tunable in both power and efficiency, with potential implications for controlling thermal transport in nanoscale systems~\cite{Butenko2014,Wang2025}.\par

\textit{Results for Interacting Gas Mixtures}---To demonstrate the universality of gravity-induced TR, we extend our analysis to interacting many-particle gas mixtures. As a prototypical example, we consider a binary gas mixture [Fig.~\ref{model}(b)], where $N$ point particles are assigned masses $m$ and $M$ with equal probability. We set $m=1$ without loss of generality, so that $M$ represents the mass ratio. The dynamics are governed by the multiparticle collision (MPC) method~\cite{M1999,Gompper2009}. The evolution proceeds in discrete time steps, each consisting of free propagation over a duration $\tau$ followed by instantaneous stochastic collisions~\cite{MPC}. Since the collision frequency scales as $\tau^{-1}$, the time step $\tau$ controls the effective interaction strength and thus the transport properties (e.g., thermal conductivity~\cite{Luo2025})). Specifically, the non-interacting limit corresponds to $\tau = \infty$, while progressively smaller $\tau$ values correspond to stronger interparticle interactions.\par

\begin{figure}\centering\includegraphics[width=8.5cm]{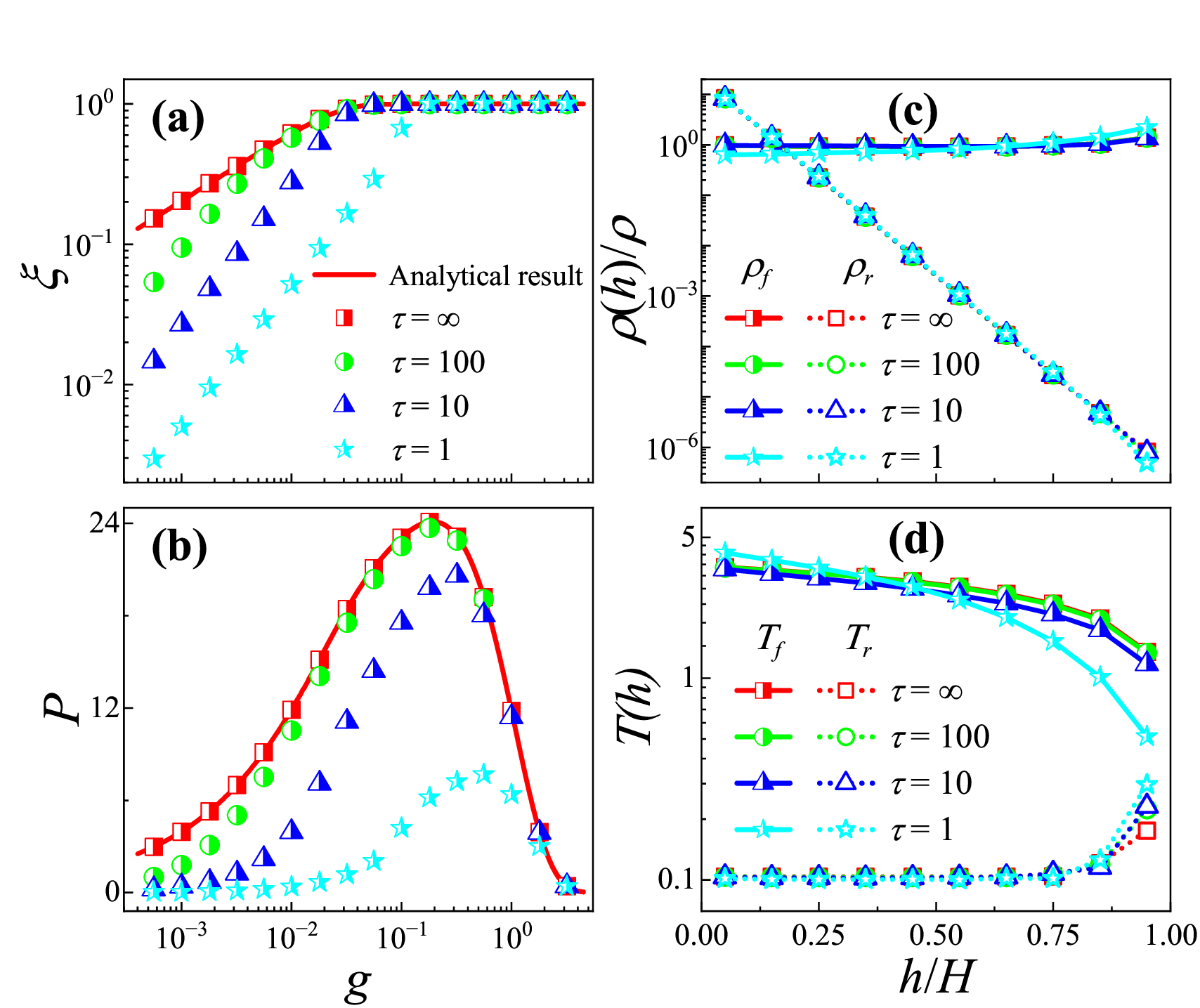}
\caption{(a) Rectification efficiency $\xi$  and (b) power $P$  versus gravitational acceleration $g$ for the 2D interacting single-component gas. Symbols correspond to different MPC time steps $\tau$; red solid lines denote the analytical non-interacting limit ($\tau = \infty$) obtained by generalizing Eq.~(\ref{xi}). Height profiles of (c) number density $\rho(h)$ and (d) local temperature $T(h)$ at $g=0.18$. Fixed parameters: $T_h=5$, $T_c=0.1$, $\theta = \pi/2$, $N =120$, $H=10$, $L=2$, $M=a=k_B=1$, and mean density $\rho=N/(LH)$.}\label{fig3}
\end{figure}

Numerical results for $\xi$ and $P$ in an interacting single-component gas ($M=1$), with $T_h=5$ and $T_c=0.1$, are presented in Figs.~\ref{fig3}(a) and~(b). Decreasing $\tau$ substantially suppresses both $\xi$ and $P$ across the rectification region. Nevertheless, TR persists over a finite range of $g$, and perfect unidirectional conduction ($\xi=1$) is attained near the maximum power $P_{\textrm{max}}$ for all $\tau$. These results demonstrate that the gravity-induced TR mechanism remains highly robust in the presence of interparticle interactions. To elucidate the characteristics of perfect unidirectional conduction, Figs.~\ref{fig3}(c) and~(d) display the height-dependent particle number density $\rho(h)$ and temperature profile $T(h)$ at $g=0.18$ (where $\xi=1$ for all $\tau$), computed using the numerical method of Ref.~\cite{Zhang2026}. In the reverse configuration, the $\rho_r(h)$ profiles for different $\tau$ values nearly coincide and exhibit an exponential decay with height, consistent with the equilibrium Boltzmann distribution; simultaneously, $T_r(h)\approx T_c=0.1$  remains essentially uniform throughout the system. This behavior arises because the average thermal energy of particles in the cold lower bath ($k_B T_c \sim 0.1$) is insufficient to overcome the gravitational potential barrier ($MgH \sim 1.8$) required to reach the upper bath. Consequently, the majority of particles accumulate near the bottom, effectively suppressing reverse heat conduction. Moreover, interparticle collisions efficiently thermalize these particles to $T_c$, yielding near-equilibrium behavior. By contrast, the forward configuration exhibits qualitatively different features: $\rho_f(h)$ approaches a spatially uniform distribution, while $T_f(h)$ develops a pronounced temperature gradient---particularly under stronger interactions. In this regime, the average energy of particles in the hot lower bath ($k_B T_h \sim 5$) exceeds the gravitational potential barrier, and interparticle interactions sustain local thermal equilibrium, thereby enabling forward heat conduction.

\begin{figure}
\centering
\includegraphics[width=8.5cm]{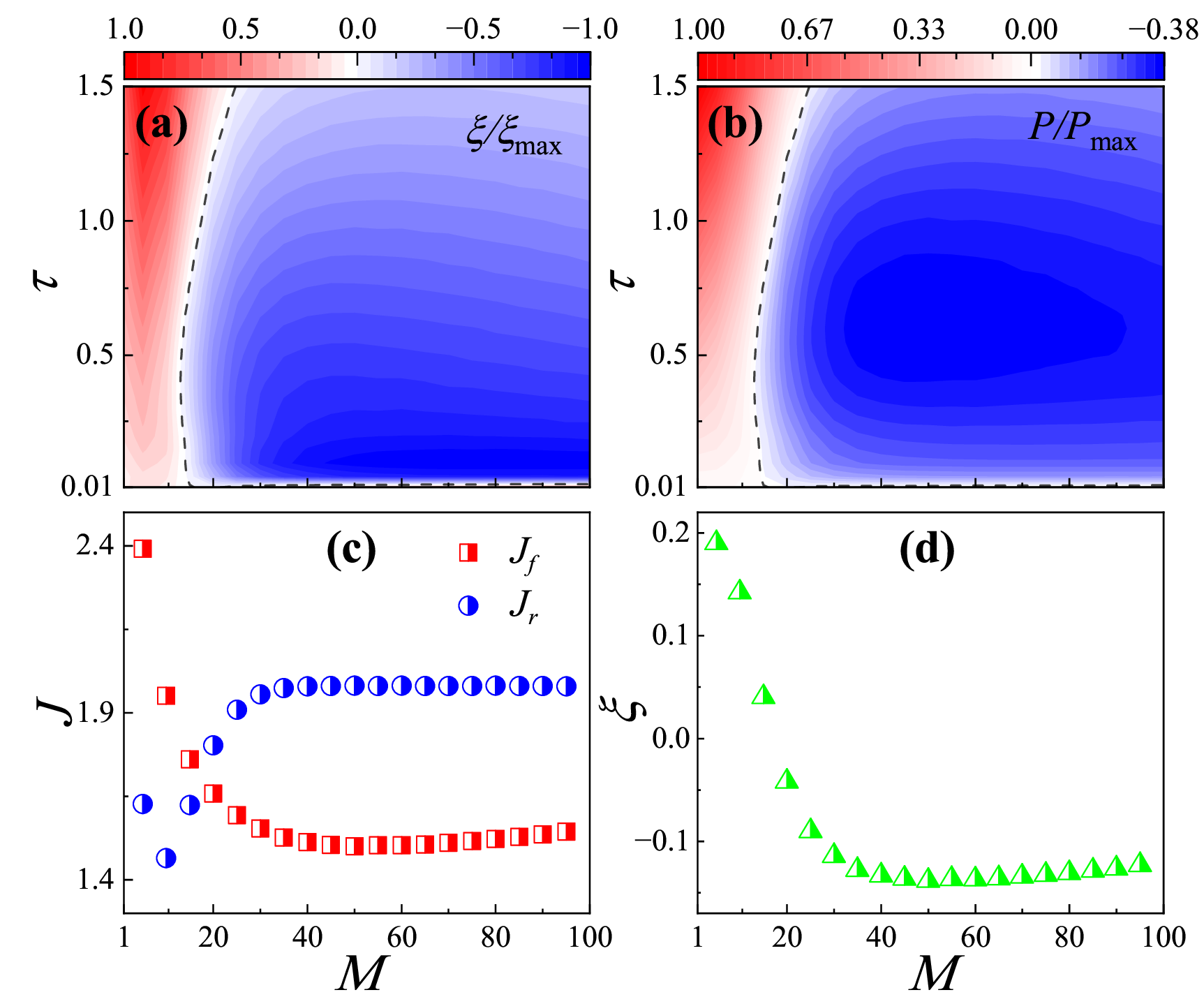}
\caption{(a) Rectification efficiency $\xi$ and (b) power $P$ as functions of interaction strength $\tau$ and binary mass ratio $M$ at $g=0.025$. Black dashed contours mark the zero-efficiency ($\xi=0$) and zero-power ($P=0$) boundaries, respectively. (c) Forward ($J_f$) and reverse ($J_r$) heat fluxes versus $M$ for $\tau=1$, traversing regions I and II. (d) Corresponding rectification efficiency $\xi$ versus $M$. All other parameters match Fig.~\ref{fig3}.}
\label{fig4}
\end{figure}

Figures~\ref{fig4}(a) and~(b) summarize $\xi$ and $P$ for binary gas mixtures as functions of the interaction strength $\tau$ and mass ratio $M$ at $g=0.025$. The $\tau$-$M$ parameter space is partitioned into two distinct regions by the critical contours $\xi=0$ and $P=0$. In region I, both $\xi$ and $P$ decrease with decreasing $\tau$, consistent with the trends observed in Fig.~\ref{fig3}. Remarkably, region II exhibits reverse rectification (indicated by the blue shading) over a broad parameter domain. As $\tau$ is reduced, $\xi$ approaches the perfect reverse limit ($\xi=-1$), while $P$ displays non-monotonic behavior---first increasing, then decreasing---revealing an inherent trade-off between maximizing rectification efficiency and power. Figs.~\ref{fig4}(c) and~(d) further characterize the $M$ dependence of the heat fluxes and $\xi$, respectively, demonstrating a clear transition from forward to reverse rectification with increasing mass ratio $M$. Although the microscopic mechanism underlying reverse TR requires further elucidation, its observation in the present work---alongside earlier reports in two-segment Frenkel-Kontorova models~\cite{Bambi2006}, anisotropic Heisenberg spin chains~\cite{Lifa2009}, and qubit-resonator systems~\cite{Paladino2025}---provides compelling evidence that reverse TR constitutes a universal phenomenon across diverse physical systems.


\textit{Conclusions and Discussion}---In summary, we have established gravity as a fundamental mechanism for inducing robust TR in simple gaseous systems, obviating the need for engineered structural asymmetry. By combining analytical theory with numerical simulations, we demonstrated that in the non-continuum regime, a single-particle gas under gravity can exhibit perfect unidirectional heat conduction over a broad parameter range, unveiling an intrinsic trade-off between rectification efficiency and power. Our numerical results further confirm the persistence of this mechanism in interacting many-particle systems. Notably, the observed rectification reversal in binary mixtures reveals a nontrivial interplay among gravity, interparticle interactions, and mass disparity, although its microscopic origin warrants further systematic investigation.\par

Given the rapid experimental progress with macromolecular and granular gases~\cite{Barmatz2007,Harth2013,Hou2008,Puzyrev2024}, the rectification mechanism elucidated in this work presents a promising candidate for experimental realization in such systems. This prospect opens a pathway toward practical applications, including gaseous thermal diodes. Beyond pure heat transport, the gravity-driven multiparticle collision framework developed here provides a computationally efficient platform for investigating coupled heat and particle transport phenomena, such as inverse currents~\cite{Wang2020} and thermoelectric effects~\cite{Luo2020,Luo2018,Benenti2014}. Extending the present study to address coupled transport in open systems represents a promising and challenging direction for future research.\par

\textit{Acknowledgments}---We acknowledge support from the National Natural Science Foundation of China (Grant No.12475034) and the Natural Science Foundation of Fujian Province (Grant No.2023J05100).

\bibliography{paper}

\begin{thebibliography}{56}
\bibitem [{sim()}]{simulation}%
  \BibitemOpen
  \href@noop {} {}\bibinfo {note} {In all simulations, we initialize the system
  with random particle position and velocity sampled from a Boltzmann
  distribution at temperature $(T_h+T_c)/2$. After the relaxation, the system
  is evolved a duration of $t=10^8$ to evaluate the heat current and collision
  frequency. The heat current is computed as the energy transfer per unit time
  between the particle and the heat bath. Since the statistical errors are less
  than $1\%$, error bars are omitted in all figures.}\BibitemShut {Stop}%
\bibitem [{MPC()}]{MPC}%
  \BibitemOpen
  \href@noop {} {}\bibinfo {note} {Within the MPC framework, the system evolves
  in discrete time steps, each consisting of a streaming phase of duration
  $\tau$, followed by instantaneous stochastic collisions. During the streaming
  step, each particle evolves under the gravitational force. For the collision
  step, the simulation domain is partitioned into square cells of side length
  $a$. The velocities of all particles within a given cell are rotated about
  their center-of-mass velocity $\mathbf{V}_{\mathrm{cm}}$ by an angle
  $\pm\theta$, chosen with equal probability. Specifically, the post-collision
  velocity of the $i$-th particle is updated as $\mathbf{v}_i' =
  \mathbf{V}_{\mathrm{cm}}+\hat{\mathcal{R}}^{\pm\theta}(\mathbf{v}_i-\mathbf{V}_{\mathrm{cm}})$,
  where $\hat{\mathcal{R}}^{\theta}$ denotes the 2D rotation operator by angle
  $\theta$. We adopt a rotation angle $\theta=\pi/2$, which maximizes momentum
  mixing and ensures optimal thermalization. These collision rules strictly
  conserve both the total energy and total momentum of the gas.}\BibitemShut
  {Stop}%
\end{thebibliography}

\end{document}